\documentclass[letterpaper, 10 pt, conference]{ieeeconf}  

\IEEEoverridecommandlockouts                              

\usepackage{upgreek}
\usepackage[utf8]{inputenc}
\usepackage{graphicx}
\usepackage{amssymb}
\usepackage{amsmath}
\usepackage{lipsum}
\usepackage{fullpage}
\usepackage{microtype}
\usepackage{lineno}
\usepackage{acronym}
\usepackage[hidelinks]{hyperref}
\usepackage{tabularx}
\usepackage{float}
\usepackage{fancyhdr}
\usepackage{caption}
\usepackage{subcaption}
\usepackage[yyyymmdd,hhmmss]{datetime}
\usepackage{chngcntr}
\counterwithout{figure}{section}
\usepackage{atbegshi,picture}
\usepackage{lipsum}

\usepackage[nospace]{cite}

\usepackage{placeins}
\usepackage{fancyhdr}
\usepackage{booktabs}
\usepackage{multirow}
\usepackage{palatino} 

\usepackage{ulem}
\usepackage{color}

\usepackage{array}
\newcolumntype{P}[1]{>{\centering\arraybackslash}p{#1}}

\newcommand{\authorlistannotated}{
  Atalanti A. Mastakouri$^{1}$,
  Bernhard Sch{\"o}lkopf$^{1}$ and
  Moritz Grosse-Wentrup$^{1,2}$
}
\newcommand{\authoraffiliations}{
  \hspace{-3.5pt}$^{1}$Max Planck Institute for Intelligent Systems, Empirical Inference department, T\"ubingen, Germany \\
  $^{2}$Research Group Neuroinformatics, Faculty of Computer Science, University of Vienna, Austria\\
}

\usepackage[shadow,linecolor=black,textsize=small]{todonotes}

\usepackage{flushend}

\title{\LARGE \bf Beta Power May Mediate the Effect of Gamma-TACS on Motor Performance}

\author{\authorlistannotated\\
\authoraffiliations}
\begin{document}

\maketitle

\thispagestyle{empty}
\pagestyle{empty}

\begin{abstract}
Transcranial alternating current stimulation (tACS) is becoming an important method in the field of motor rehabilitation because of its ability to non-invasively influence ongoing brain oscillations at arbitrary frequencies. However, substantial variations in its effect across individuals are reported, making tACS a currently unreliable treatment tool. One reason for this variability is the lack of knowledge about the exact way tACS entrains and interacts with ongoing brain oscillations. The present crossover stimulation study on 20 healthy subjects contributes to the understanding of cross-frequency effects of gamma (70 Hz) tACS over the contralateral motor cortex by providing empirical evidence which is consistent with a role of low- (12~-20 Hz) and high- (20-~30 Hz) beta power as a mediator of gamma-tACS on motor performance. 
\end{abstract}

\section{INTRODUCTION}

Transcranial alternating current stimulation (tACS) modulates neural activity and behaviour through the creation of an electric field inside the brain \cite{Herrmann2013, BESTMANN2017R1258}. More specifically, tACS applies a weak electrical alternating current on the scalp \cite{Paulus} and changes the membrane potential of the affected neurons \cite{Antal}. TACS has been used broadly in behavioural studies \cite{LOPEZALONSO2014372, Strube} as well as for the treatment of neurological disorders \cite{Schulz, Fregni1614}, although its exact neurophysiological effect on brain networks has not yet been fully understood \cite{10.3389/fnhum.2018.00211}.

TACS studies targeting motor cortex have reported considerable variability in stimulation response across individual subjects, with large percentages of non-responders \cite{LOPEZALONSO2014372, Strube}. Although tACS in the $\gamma$- (${\sim70}$ Hz) and $\beta$- (${\sim20}$ Hz) range has been proposed to facilitate and inhibit movement, respectively \cite{beta, Pogosyan20091637, TJP:TJP4253,WACH20131}, contradictory outcomes have been reported regarding the significance of these effects \cite{Moisa12053, HBM:HBM20056, Antal200897}. Much light has been shed on the role of physiological $\gamma$- \cite{gammaReview, Muthukumaraswamy2010} and $\beta$-oscillations in movement \cite{Espenhahn, GULBERTI2015, McAllister2013}. However, the manner that tACS entrains these ongoing brain oscillations is still not fully understood \cite{10.3389/fnsys.2013.00076, HELFRICH201676}.  

A short overview of the role of physiological $\beta$- and $\gamma$-oscillations in movement is important for the construction of our main argument. Activity in the $\gamma$-band has been associated both with cued and self-paced transient finger movements \cite{Muthukumaraswamy2010}. Furthermore, relatively large ballistic movements of greater movement amplitude were associated with increased motor cortex $\gamma$-power \cite{Muthukumaraswamy2010}. Moreover, Gaets et al.~gave evidence for a motor $\gamma$-band network for response selection and maintenance of planned behaviour \cite{GAETZ2013245}. These observations justify our selection of $70$ Hz for stimulation of the contralateral motor cortex in the present study.

On the other hand, $\beta$-oscillatory activity has been found to be significantly elevated in patients with motor disorders (tremors, slowed movements, trouble initiating movements) such as Parkinson’s disease  \cite{McAllister2013, BROWN2007656, Khanna2017}. Furthermore, for healthy subjects, it was reported that movements preceded by a reduction in $\beta$-power exhibited significantly faster reaction times than movements preceded by an increase in $\beta$-power \cite{Khanna2017}. It has also been proposed that $\beta$-activity represents the status quo \cite{ENGEL2010156}, suggesting that enhanced $\beta$-activity prevents change from the current state \cite{Schnitzler2005, Davis403}.

Based on existing knowledge about the role of $\beta$-oscillations in the inhibition of movement speed \cite{McAllister2013}, and about the effect of high stimulation frequencies on the decrease of $\beta$-power \cite{GULBERTI2015}, we hypothesize that the modulation of the ongoing $\beta$-activity mediates the effect of $\gamma$-tACS on the behavioural response to the stimulation. This hypothesis has two empirically testable implications: First, we expect arm speed to be affected by stimulation. We examine the effect of $\gamma$-tACS on movement in Section \ref{test1Methods}. Seccond, we expect $\beta$-power to be affected by the stimulation. For that purpose, we examine if this is true and, if so, in which brain areas a modulation of $\beta$-power can be observed (cf.~Section \ref{test2Method}). We expect that changes in $\beta$-power induced by $\gamma$-tACS are only significant in the subgroup of subjects that exhibit a behavioural response to the stimulation. While these two empirical observations would be consistent with a role of $\beta$-power as mediator of the effect of $\gamma$-tACS on behavioural performance, causal relations between brain oscillations can not be substantiated by correlational evidence only. We thus perform an additional causal analysis, based on the Information Geometric Causal Inference algorithm \cite{Daniusis2010}, applied here for the first time to EEG data, examining the effect of $\beta$-power change on motor performance (Section \ref{test3Method}). The results of this test lend further support to a potentially causal role of $\beta$-power in the response to $\gamma$-tACS. We discuss the neurophysiological plausibility of our findings in Section \ref{neurophysiologicalExplanation}.

 
The study conformed to the Declaration of Helsinki, and the experimental procedures involving human subjects described in this paper were approved by the Ethics Committee of the Medical Faculty of the Eberhard Karls University of Tübingen.

\section{METHODS}
\subsection{Experimental Setup}

\subsubsection*{Subjects}
Twenty healthy, right-handed subjects participated in this study. One of the subjects (ID 10) did not participate in the second day of the recordings and was excluded from the analysis. The remaining 19 healthy participants (nine female, ten male) were $28.36 \pm 8.57$ years old.

\subsubsection*{Stimulation parameters}
We chose a crossover design in which both real- and sham stimulation are applied to each subject in a randomised order. High-definition-tACS (HD-tACS) was used for the stimulation (DC Stimulator Plus, Neuroconn). The HD 4$\times$1 setup was preferred over the common two-electrode setup in order to increase the focality of the stimulation on the preferred motor area \cite{Dmochowski}. The equalizer extension box was used to extend the two ordinary square sponge electrodes into a 4$\times$1 set of round rubber electrodes of $20$ mm diameter, with one anode on the region to be stimulated and four cathodes in a square around it, each cathode at $~7.5$ cm from the central anode \cite{Villamar}. The anode was placed on channel C3 (primary motor cortex -- M1) and the four cathodes on Cz, F3, T7, and P3, following the instructions described in \cite{Villamar}. For both, real- and sham blocks, a duration of $15$ min was chosen \cite{Struber}.

For the real stimulation, a sinusoidal signal mode at $70$ Hz with a peak-to-peak amplitude of $1$ mA was used. For the sham stimulation, a sinusoidal signal at $85$ Hz with a peak-to-peak amplitude of $1$uA was selected. 
\begin{figure*}[thpb]
\center
\includegraphics[width=\textwidth]{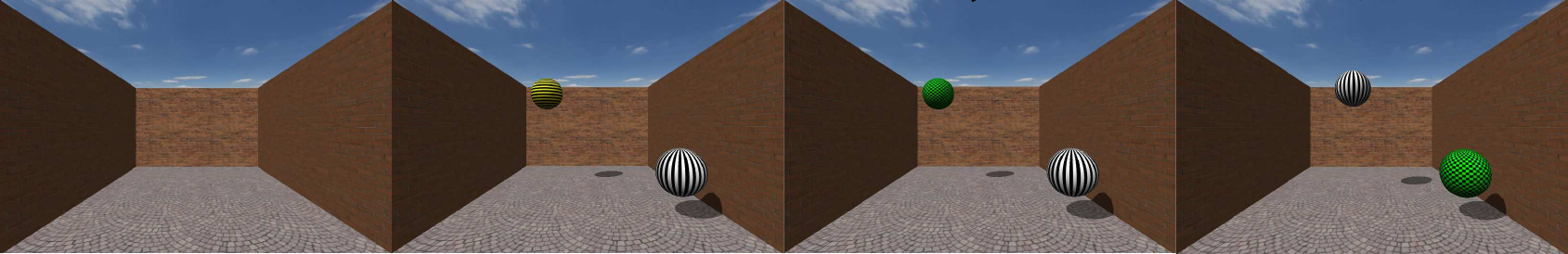}
\caption{Paradigm: The white sphere represents the real-time position of the subject's right arm. The yellow/green sphere represents the target, appearing at a random location in each trial. Participants were instructed to reach for the target when its color changed from yellow to green. After the subject successfully reached the target, a green sphere appeared at the starting position, indicating to return their hand to the starting position to complete the trial.\label{fig:setup}}
\end{figure*}
\subsubsection*{Paradigm}
Each participant attended two sessions, separated by a one-day break. On the first session (day 1), participants performed a visuomotor target-reaching task with their right arm, consisting of three blocks of 50 trials each, while their brain activity was recorded with EEG. Each block was separated by a $5$-minutes resting-state period, during which the participant was asked to relax and focus on a white cross on the black screen in front of them.
On the second session (day 2), either real HD-tACS or sham-stimulation was applied in a randomised order (blinded to the subject) during the second and third block, respectively. The order of real-/sham stimulation in the second and third block was randomised across subjects in order to compensate for unknown factors such as learning effects or tiredness over time. The second session consisted of three blocks, with a $20$-minute break between the second and third block to avoid carry-over effects between the two blocks. Each block consisted of as many random reaching--trials as the subject could complete in $15$ minutes.

The participants were seated on a chair in the middle of four infrared motion tracking cameras (PhaseSpace), facing a screen and wearing a specially designed glove with three LEDs for real-time tracking of their arm position, which was depicted in real-time as a 3D sphere, as shown in Fig.~\ref{fig:setup}.
\subsubsection*{Experimental data}
The experimental data include motion tracking data of the subjects' arm position, recorded with $f_s=960$ Hz, and EEG data from high-density EEG (128 channels, $f_s=500$ Hz, Brain Products GmbH).  
\subsection{Motor Response to $\gamma$-tACS}
For the analysis in the present paper, we focused on the data of the stimulation day (day 2). For each subject and trial, we computed the mean movement velocity. The trials of each block, in which movement speed exceeded three standard deviations, were excluded as outliers. 
\subsubsection*{Division into responders and non-responders based on movement velocity} 
Based on the movement response of each subject to stimulation, we categorised subjects into two groups. For each subject, we tested the null hypothesis that the average movement velocity over the stimulation block is the same as the movement velocity during the sham block. For each subject, we performed a permutation t-test. To build the null-distribution, we concatenated the velocities of the two blocks and permuted them $10^4$ times, computing the mean velocity of each of the two blocks after every permutation. We calculated the $p$-value as the frequency at which we found the absolute difference between mean velocity during real- and sham stimulation not to be larger than when drawing from the null-distribution (two-sided test). Setting a threshold $\alpha = 0.05$, we categorised subjects into two groups: \textit{Responders} if $p < \alpha$ and the average movement velocity during the stimulation block was greater than during the the sham block, and \textit{ non-responders} otherwise, i.e., subjects who did not show a significant increase or who show a decrease in movement speed in response to stimulation.\label{test1Methods} 
\subsubsection*{Effect size}
We quantified the effect size of $\gamma$-tACS over contralateral M1 for each subject as the difference between the average arm speed during the stimulation and during the sham block, normalized by the standard deviation during the sham block.

\subsection{Effect of $\gamma$-tACS on $\beta$-power}
To attenuate non-cortical artifacts in the EEG data, we concatenated the EEG signal of each subject's resting states between blocks, high-pass filtered the data with a Butterworth filter with cut-off frequency at $3$Hz, applied a common-average reference filter, and then performed SOBI Independent Component Analysis (ICA) followed by manual rejection of non-cortical sources \cite{MCMENAMIN20102416}.

To examine the effect of $\gamma$-stimulation on $\beta$-activity, and the relation of $\beta$-power with motor performance, we calculated the log-bandpower of the 116 z-scored channels for each subjects (after having removed the channels used for stimulation) for the low- (12--20) Hz and high- (20--30) Hz $\beta$-range. For visualization purposes, we calculated the group grand-averages of the difference between $\beta$-log-bandpower after and before the stimulation for the stimulation- as well as for the sham block. 

To test within each group (\textit{responders} and \textit{non-responders}) whether the changes observed in $\beta$-power are statistically significant, we performed a permutation paired t-test with $10^4$ permutations (two-sided): For each channel, we tested the null hypothesis that the neurophysiological changes in $\beta$-log-bandpower during stimulation come from the same distribution (across subjects) as those during sham. We performed FDR-correction for multiple testing at significance level $\alpha = 0.05$ \cite{fdrBenjaminiHochberg}\label{test2Method}. 


\subsection{Causal analysis of $\beta$-power and movement speed}

To directly test for causal relationships between neurophysiological changes in $\beta$-power and motor performance, we applied a causal inference analysis on the channels of each group. More specifically, we examined which channels express the causal relationship $\Updelta\beta_{\text{BP}_\text{{stimulation}}} \rightarrow \text{Effect size}$, where $\Updelta\beta_{\text{BP}_\text{{stimulation}}} = \beta_{\text{BP}_\text{{after stimulation}}} - \beta_{\text{BP}_\text{{before stimulation}}}$. To do so, we used the \textit{Information Geometric Causal Inference} (IGCI) algorithm proposed by \cite{Daniusis2010}, which is applied here for the first time to study causal relationships between brain oscillations and behaviour. The IGCI inference algorithms is based on the assumption that if $X \rightarrow Y$, the distribution of $X$ and the function $f$ that maps $X$ to $Y$ are independent, i.e., it assumes that the mechanism and the data that it processes are not co-adapted. Independence between the function $f$ and the distribution of $X$ is computed by the relative entropy distance $D(.,.)$, which is then used to estimate $C_{X \rightarrow Y} = D(p_{X}||E_{X}) - D(p_{Y}||E_{Y})$. Based on the sign of $C_{X \rightarrow Y}$, the IGCI algorithm decides which causal direction is more likely. If $C_{X \rightarrow Y}>0$, then $X$ is inferred as the cause of $Y$. We note that this test does not take into account the possibility of hidden confounders.\label{test3Method}

\section{RESULTS}

\subsection{Motor response to $\gamma$-tACS}
\begin{figure*}[thpb]
\center
\includegraphics[width=0.9\textwidth]{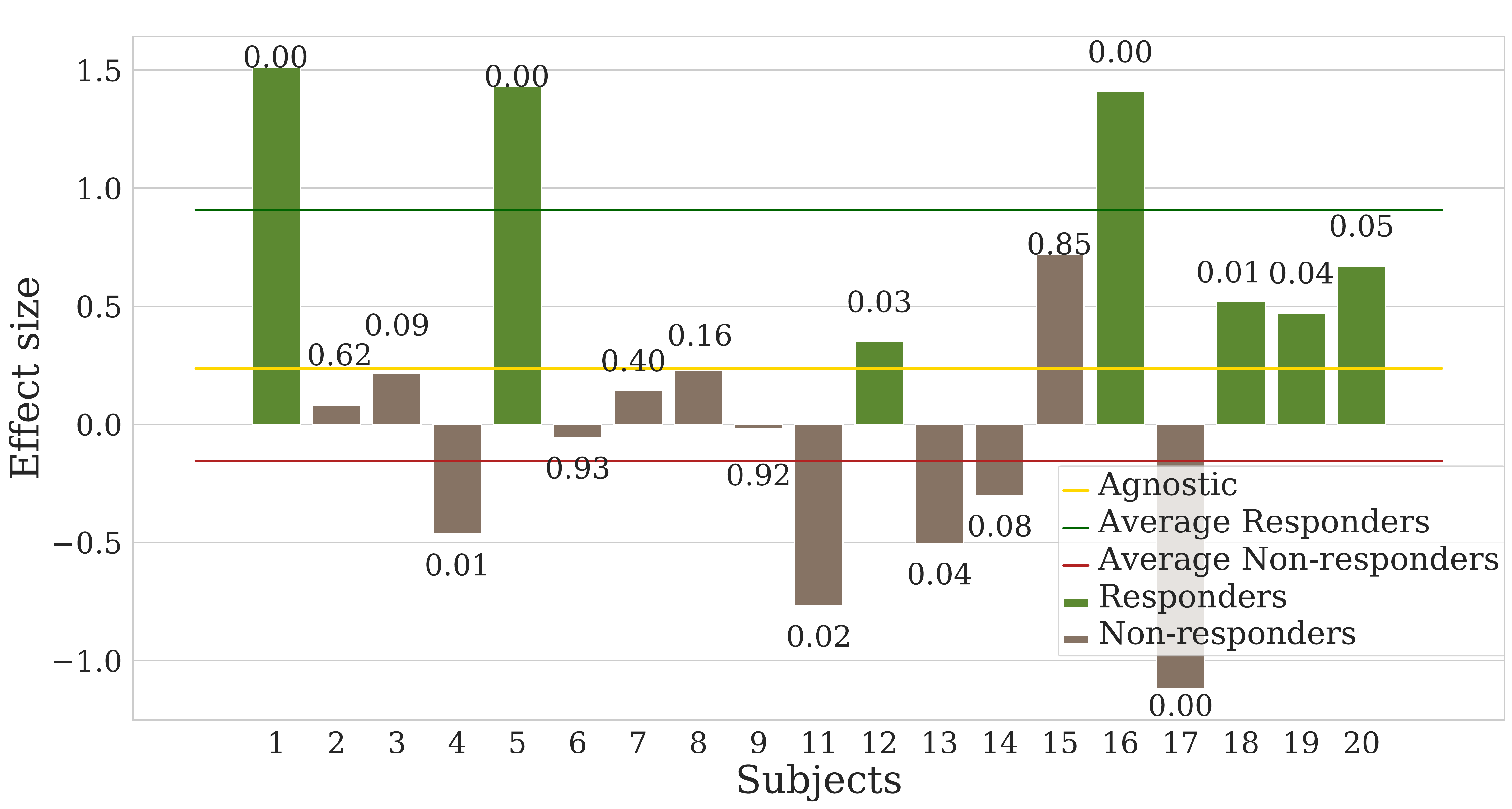}
\caption{Individual effect sizes for stimulation as measured by changes in movement velocity. The $p$-value (rounded up to two decimals) for each subject is shown on top of each bar (cf.~\ref{test1Methods}). Green bars: Subjects that performed significantly better during stimulation (responders). Brown bars: Subjects who either did not respond to the tACS, or who performed significantly worse compared to sham (non-responders). Green line: Average effect size of responders. Red line: Average effect size of non-responders. Yellow line: Overall effect size of the whole population.\label{fig:indivEffectSizes}}
\end{figure*}

Out of the original population of 19 subjects, only six subjects responded positively to $\gamma$-tACS over contralateral M1. Figure~ \ref{fig:indivEffectSizes} depicts the effect size for each subject, with the color indicating whether the subject was a responder or a non-responder. Contralateral M1 $\gamma$-tACS has been proposed in several studies as a stimulation setup that facilitates movement. Here, however, the overall effect size is quite small ($0.2366$) due to the existence of responders and non-responders with effect sizes $0.9073$ and $-0.1547$, respectively. Based on these observations, we next examined the effect of $\gamma$-tACS on $\beta$-powerfor each of the two groups individually.\label{Test1Res}

%
%
%
%

\subsection{Effect of $\gamma$-tACS on $\beta$-power}

\begin{figure*}[thpb]
\center
\includegraphics[width=\textwidth]{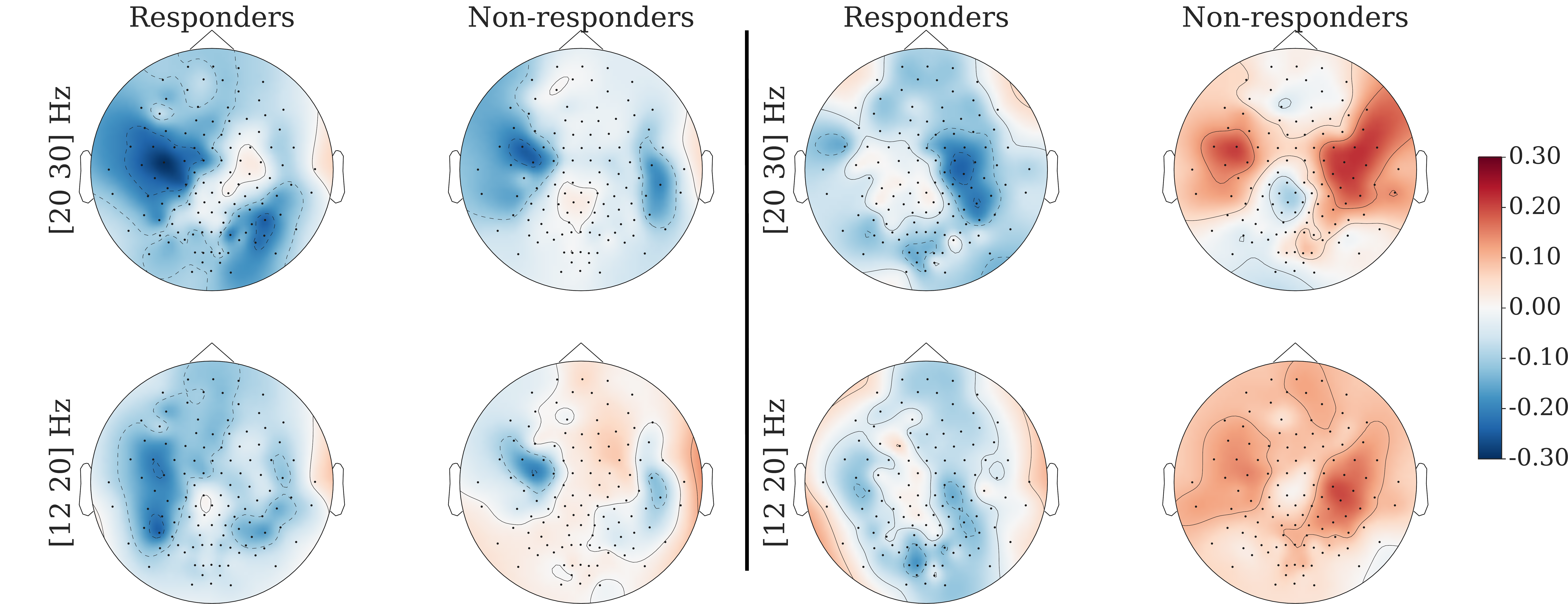}
\caption{Difference between $\beta$-power after and before stimulation (left) and sham (right), in the low- (12--20) Hz and high- 20--30 Hz $\beta$-range for the groups of responders and non-responders.\label{fig:diffSham_better_vs_worse}}
\end{figure*}

We hypothesized that subjects, who exhibit a larger decrease in $\beta$-log-bandpower over the contralateral motor cortex, are those that respond positively to the stimulation, i.e., with faster movements. In Figure \ref{fig:diffSham_better_vs_worse}, we see that the group of subjects, that responded positively to the $\gamma$-stimulation, show a larger decrease of $\beta$-power, mostly in the high $\beta$-range $[20 ~30]$ Hz and spread out over the contralateral motor cortex, compared to the group of the subjects that did not respond to stimulation.\label{Test2Res}

Moreover, for the sham condition in the group of responders, we found little change of $\beta$-log-bandpower over contralateral motor cortex. In contrast, the group of non-responders exhibits a bilateral increase of $\beta$-power.

\begin{figure*}[thpb]
\center
\includegraphics[width=0.95\textwidth]{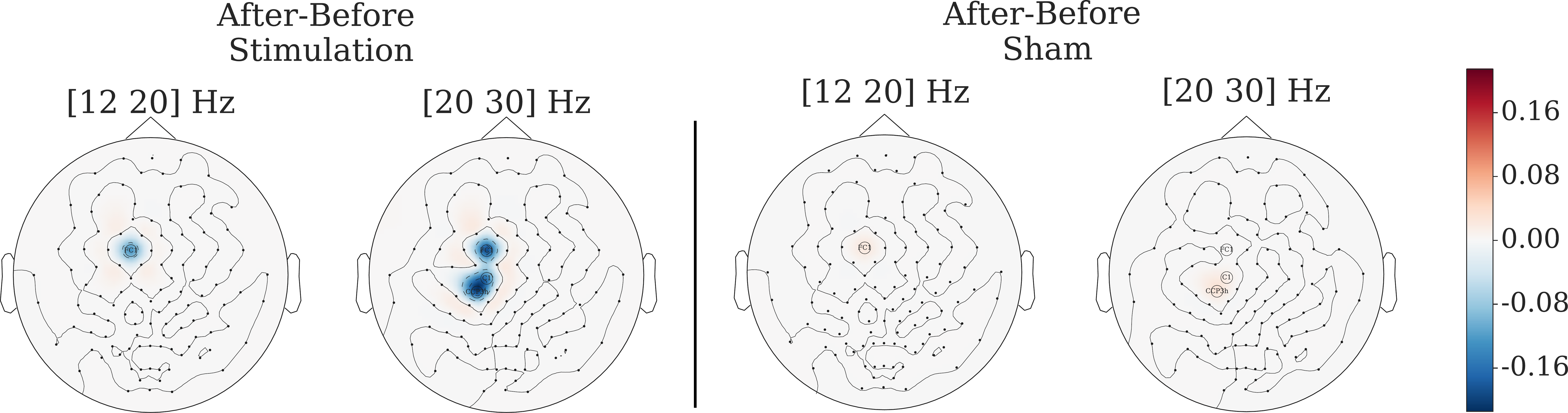}
\caption{Difference of $\beta$-power after and before stimulation for the real- and sham stimulation in the low (12--20 Hz) and high- (20--30 Hz) $\beta$-range, for the group of responders. FDR-corrected channels, that do not exhibit significant neurophysiological differences, are set to zero.\label{fig:fdr_chans_diffStimDiffSham}}
\end{figure*}

Figure \ref{fig:fdr_chans_diffStimDiffSham} depicts the channels that exhibit FDR-corrected statistical significance of differences in $\beta$-power, between the conditions of real- and sham stimulation, for each group. For the responders group, these channels are found to be located over the contralateral motor cortex, FC1, C1 and CCP3h for the high $\beta$-range and FC1 for the low $\beta$-range. For the non-responders, in contrast,  we found no channel with a significant difference between the two conditions.

\subsection{Causal analysis of $\beta$-power and movement speed}

\begin{figure*}[thpb]
\begin{center}\includegraphics[width=0.95\textwidth]{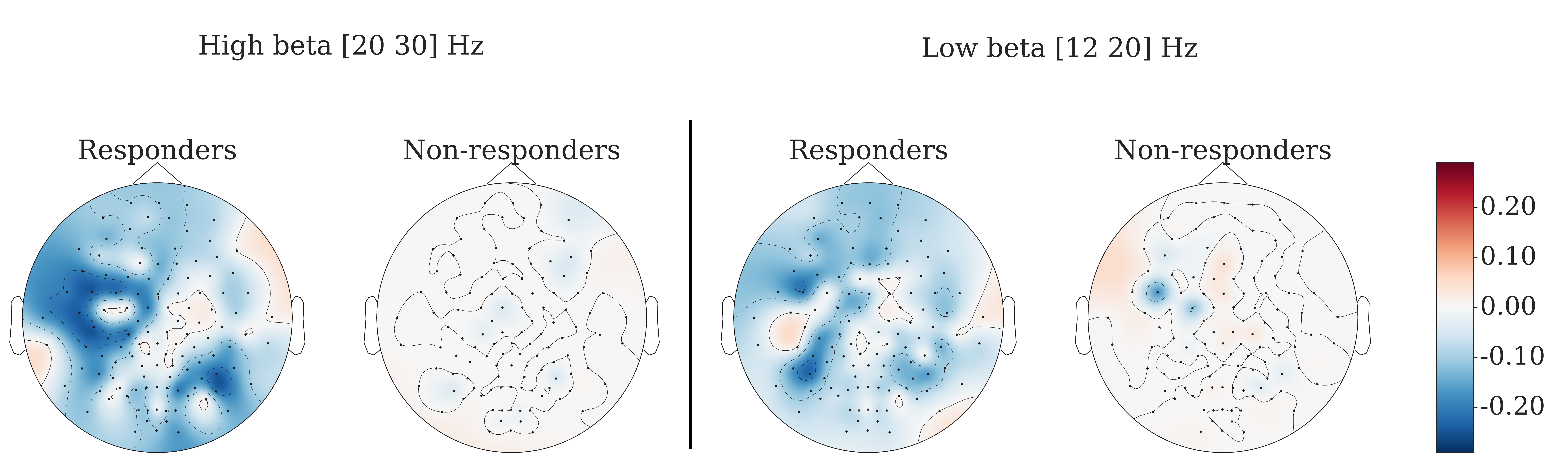}
\end{center}
\caption{Difference of $\beta$-power after and before stimulation, in the low- (left) and high (right) $\beta$-range, with channels that do not satisfy the causal relationship $\Updelta\beta_{\text{BP}_\text{{stimulation}}} \rightarrow \text{Effect size}$ set to zero.\label{fig:causalChannelsDiffBetaEffectSize}}
\end{figure*}

The causal inference analysis, for the detection of channels that satisfy the relationship $\Updelta\beta_{\text{BP}_\text{{stimulation}}} \rightarrow \text{Effect size}$, was applied independently for each of the two groups as well as for the low- and high $\beta$-range. In the low $\beta$-range, no channel was found to satisfy the above relationship. For the high $\beta$-range, the channels that did not satisfy the condition were set to zero. The remaining channels are depicted in Fig.~\ref{fig:causalChannelsDiffBetaEffectSize}, colour-coded according to the difference between high $\beta$-log-bandpower after and before stimulation. We observe that for the responders group the left motor cortex exhibits the above causal relationship. In contrast, the majority of the channels for the group of non-responders do not satisfy the causal relationship.\label{Test3Res} 

\section{DISCUSSION}


Applying HD-tACS at $70$ Hz over contralateral motor cortex on 20 healthy subjects, we found a significant increase of upper-limb movement speed in $36\%$ of the original population. Consistent with the results in \cite{LOPEZALONSO2014372}, we found a large number of non-responders. Considering the fact that $\gamma$-stimulation is believed to facilitate movement \cite{beta,LOPEZALONSO2014372}, as well as that an increase of $\gamma$-oscillatory activity has been associated with large ballistic movements \cite{Muthukumaraswamy2010}, we decided to investigate the underlying modulation of the antikinetic $\beta$-oscillations \cite{Brinkman14783} as a possible cause of this variability. We hypothesized that if $\gamma$-stimulation is affecting the ongoing $\beta$-oscillations, then the subjects that exhibit a larger decrease in $\beta$-power should be those that respond to the stimulation. Our EEG findings support this hypothesis.\\

Taken together, our findings support a potential role of $\beta$-power as a mediator of $\gamma$-tACS on motor performance. In particular, the results reported in Section \ref{Test1Res} establish that $\gamma$-tACS ($S$) has an effect on movement performance $(P)$, as measured by movement velocity. Because $S$ is a randomised treatment, we can infer the direction of this relation as $S \rightarrow P$. The results in Section \ref{Test2Res}, on the other hand, demonstrate an effect of $\gamma$-tACS on $\beta$-power, i.e., a causal path $S \rightarrow \Updelta\beta$. It then remains to distinguish between the two causal models $S \rightarrow \Updelta\beta \rightarrow P$ (with potentially an additional path $S \rightarrow P$ that does not pass through $\Updelta\beta$) and $\Updelta\beta \leftarrow S \rightarrow P$, both of which are consistent with our evidence to this point. The results of the causal analysis in Section \ref{Test3Res} indicate that in the stimulation condition $\Updelta\beta \rightarrow P$, which is consistent with the former and not with the latter causal model. We thus argue that our empirical results are in favour of the causal model $S \rightarrow \Updelta\beta \rightarrow P$, i.e., that $\beta$-power may mediate the effect of $\gamma$-tACS on motor performance. We stress, however, that causal inference methods as applied here can not prove but only provide empirical results consistent with causal relationships.\\  

\label{neurophysiologicalExplanation}We further note, however, that the causal model $S \rightarrow \Updelta\beta \rightarrow P$ is neurophysiologically plausible. In the context of neurophysiological procedures underlying the effect of $\gamma$-stimulation on $\beta$-power and on the observed motor behaviour, a possible explanatory factor could be the modulation of $\gamma$-aminobutyric acid (GABA) concentration. We support this claim with the following argument. First, $\beta$-oscillations have been shown to be the summed output of principal cells temporally aligned by GABAergic interneuron rhythmicity \cite{YAMAWAKI2008386}. Specifically, GABA levels have been found to strongly correlate with $\beta$-power and to exhibit elevated values in bradykinesia and in Parkinson's disease \cite{McAllister2013}. Secondly, high-$\gamma$ deep brain stimulation in motor cortex has been reported to cause a significant decrease in $\beta$-power \cite{GULBERTI2015}, supporting our finding of the inhibitory effect of $\gamma$-stimulation on the ongoing $\beta$-oscillations. Combining these two literature research conclusions, we argue that the behavioural response to $\gamma$-tACS may be explained by a decrease of $\beta$-power and hence of GABA levels, modulated by the stimulation. We note that it is also conceivable that whenever $\gamma$-tACS leads to the inhibition of human movements, this may be caused by an increase in GABAergic drive, which hinders $\beta$-power to be decreased.

\bibliographystyle{IEEEtran}

\bibliography{mybibfile}





\end{document}